\pdfoutput=1
\documentclass[aps,prx,twocolumn,floatfix]{revtex4-2}

\usepackage{bm}
\usepackage{amsfonts}
\usepackage[pdftex]{graphicx}
\usepackage[usenames,dvipsnames]{color}
\usepackage{amssymb}
\usepackage{amsmath}
\usepackage{bbold}
\usepackage{algorithm}
\usepackage{algpseudocode}

\usepackage{longtable}
\usepackage{array}
\usepackage{dcolumn}
\newcolumntype{P}[1]{>{\centering\arraybackslash}p{#1}}
\newcolumntype{M}[1]{>{\centering\arraybackslash}m{#1}}
\newcolumntype{B}[1]{>{\centering\arraybackslash}b{#1}}

\usepackage{bbm}
\usepackage{amsthm}
\usepackage{hyperref}
\hypersetup{
    colorlinks=false,
}
\usepackage{times}

\usepackage[english]{babel}
\usepackage{blindtext}
\usepackage{subfigure}

\newcommand{\ket}[1]{\left| #1 \right>}

\newcommand{\fcite}[1]{\ensuremath{^{[{\color{red}x}]}}}
\let\oldmarginpar\marginpar
\renewcommand\marginpar[1]{\-\oldmarginpar[\raggedleft\tiny\color{red} #1]
{\raggedright\tiny #1}}

\DeclareMathAlphabet\mathbfcal{OMS}{cmsy}{b}{n}
\graphicspath{{Figures/}}

\begin{document}

\title{Scalable Neural Decoder for Topological Surface Codes}
\date{\today}

\author{Kai Meinerz}
\author{Chae-Yeun Park}
\author{Simon Trebst}
\affiliation{Institute for Theoretical Physics, University of Cologne, 50937 Cologne, Germany}


\begin{abstract}
With the advent of noisy intermediate-scale quantum (NISQ) devices, practical quantum computing has seemingly come into reach. 
However, to go beyond proof-of-principle calculations, the current processing architectures will need to scale up to larger quantum circuits
which in turn will require fast and scalable algorithms for quantum error correction.
Here we present a neural network based decoder that, for a family of stabilizer codes subject to depolarizing noise
and syndrome measurement errors, 
 is scalable to tens of thousands of qubits (in contrast to other recent machine learning inspired decoders)
and exhibits faster decoding times than the state-of-the-art union find decoder for a wide range of error rates (down to 1\%). 
The key innovation is to autodecode error syndromes on small scales by shifting a preprocessing window over the underlying code, akin to a convolutional neural network in pattern recognition approaches. 
We show that such a preprocessing step allows to effectively reduce the error rate by up to two orders of magnitude in practical applications
and, by detecting correlation effects, shifts the actual error threshold
up to fifteen percent higher than the threshold of conventional error correction algorithms such as union find or minimum weight perfect matching, even in the presence of measurement errors.
An in-situ implementation of such machine learning-assisted quantum error correction will be a decisive step to push the entanglement frontier
beyond the NISQ horizon.
\end{abstract}

\maketitle


\paragraph*{Introduction.--}
In quantum computing, recent years have seen a paradigm shift which has pivoted experimental roadmaps from building
devices of a few pristine qubits towards the realization of circuit architectures of 50-100 qubits but tolerating a significant 
level of imperfections -- the advent of what has been termed noisy intermediate-scale quantum (NISQ) technology \cite{Preskill2018NISQ}. 
This move has enabled a fundamental success in the recent demonstration that such a NISQ quantum processor 
is capable of exhibiting a true `quantum advantage' over classical computing resources \cite{QuantumSupremacy}.
One of the leading NISQ platforms involves arrays of superconducting charge qubits, so-called transmons \cite{PhysRevA.76.042319},  
which by design are particular resilient with regard to charge fluctuations. However, building larger quantum circuits from transmons
comes with some intricate challenges \cite{Willsch2017Gate,Berke2020} and 
will eventually mandate to incorporate quantum error correction (QEC) schemes \cite{GottesmanErrorCorrection}. 
Arguably the most promising approach here is the implementation of a surface code \cite{KITAEV20032,bravyi1998quantum}, 
which exploits topological properties of the system
and, at the same time, remains experimentally feasible \cite{Fowler2012Surface,Wallraff_s7_2020}.
In practical settings, one downside of realizing such surface code architectures 
is the relatively slow decoding time of current quantum error correction codes.

The decoding step in quantum error correcting codes requires, at its core, a {\sl classical} algorithm that efficiently infers 
the locations of errors from measured error syndromes \cite{gottesman1997thesis}.
The most widely adopted algorithm for this purpose is minimum weight perfect matching (MWPM)~\cite{MWPM}, an algorithm which runs 
in polynomial time and is known to nearly achieve the optimal threshold for the independent noise model~\cite{Dennis2002,harrington2004analysis} (a characteristic which does not hold for more general noise models, though).
One of the drawbacks of the MWPM algorithm, however, is that its implementations are often simply too slow 
\footnote{An impressive optimization of an MWPM implementation has recently been reported 
using a local variant of the matching decoder \cite{Higgott2020Subsystem} in form of the PyMatching package \cite{PyMatching}, 
which we use for benchmarking purposes throughout this manuscript.}
even for the current generation of superconducting qubits. 
To improve the algorithmic scaling and to push error thresholds for more general noise situations, 
a number of alternative decoding approaches have been suggested, of which the most notable might be the renormalization group (RG)
\cite{RG1, RG2, Poulin2014}  and union-find (UF) \cite{UnionFind2017} decoders.
The RG decoder runs, for a surface code in a two-dimensional geometry of linear size $L$, in $O(L^2 \log L)$ time, 
often a significant improvement over the MWPM approach (which, in the worst case, scales cubic in the number of errors and thus $O(L^6)$ in code distance). 
However, its threshold value of $\sim 0.129$ for depolarizing noise \cite{RG1} is lower than that of the MWPM algorithm ($\sim 0.151$~\cite{harrington2004analysis}). 
The most efficient conventional algorithm is the recently developed UF decoder which runs in $O(L^2 \alpha(L^2))$, 
i.e. almost linear in the number of qubits 
\footnote{ The additional prefactor $\alpha(L^2)$ is the inverse of Ackermann's function whose value is $<3$ for all practical purposes.},
with a threshold $\sim 0.146$ for the depolarizing noise model (see below)~\footnote{
It has been claimed that parallelized versions of the MWPM~\cite{fowler2013minimum} or RG decoders~\cite{RG1} can achieve faster than linear decoding time. However, algorithmic scaling of such parallelized algorithms has been discussed in mostly theoretical terms and not demonstrated under realistic conditions yet (e.g. by measuring wall-clock times as done in this manuscript). In particular, we believe that resolving data dependencies will remain a significant bottleneck in any {\em practical implementation} of the proposed parallelized algorithms.}.
In addition, the last two years have seen a flurry of activity to adopt machine learning (ML) techniques to best the decoding times and threshold values of these `conventional' algorithms \cite{Varsamopoulos_2017, Torlai2017Neural, Krastanov2017,Chamberland_2018, Baireuther_2019, Liu2019Neural, Andreasson2019quantumerror, Wagner2020Symmetries, Fitzek2020Deep, Domingo2020Reinforcement, Ni2020neuralnetwork, PhysRevA.102.032411, Sweke2020Reinforcement, delfosse2020hardware}. 
As ML methods can be easily parallelized and generally offer a high degree of adaptability, one might easily accept their potential, 
but the first practical ML-based decoders have typically delivered only on one of the two benchmarks -- improving the error threshold
at the expense of scalability or the other way round, providing good scalability but leading to error thresholds which are sometimes 
even {\sl below} those of the conventional algorithms outlined above 
\footnote{We provide a snapshot of the current status of ML-based decoders in the appendix Sec. \ref{App:Comparision}
, which provides detailed performance benchmarks and allows for direct comparisons of various approaches.}.

\begin{table}[t]
\begin{tabular}{ |c||c|c|c|c|c|  }
  \multicolumn{6}{c}{depolarizing noise $(L=255)$}\\
\hline
  algorithm & $p_{th}$& $t_{p = 0.01}$ & $t_{p = 0.05}$ & $t_{p = 0.1}$ & $t_{p = 0.1461}$ \\
 \hline
 ML(7) + UF &  {\bf 0.167(0)}  & 10.5 & 25.1 & 43.4 & 78.6\\
 ML(5) + UF &  0.162(5)  & {\bf 6.7} & {\bf 12.8} & {\bf 26.2} & {\bf 56.2} \\
 Lazy + UF & 0.131(9) & 6.9 & 20.7 & 51.1 & --- \\
 UF &   0.146(1)  & 8.4 & 22.5 & 44.9 & 92.8\\
 \hline
 ML(7) + MWPM & 0.167(1)  & $\sim$ 210 & $\sim$ 530 & $\sim$ 650 & $\sim$ 980 \\
 ML(5) + MWPM & 0.163(8) & $\sim$ 270 & $\sim$ 510 & $\sim$ 650 & $\sim$ 970 \\
 MWPM   & 0.154(2)  & $\sim$ 560  & $\sim$ 840 & $\sim$ 1100 & $\sim$ 1300\\
\hline 
\end{tabular}

\vskip 1mm

\begin{tabular}{ |c||c|c|c|c|c|  }
 \multicolumn{6}{c}{ depolarizing noise + syndrome errors $(L=31)$} \\
\hline
   algorithm & $p_{th}$& $t_{p = 0.01}$ & $t_{p = 0.02}$ & $t_{p = 0.03}$ & $t_{p = 0.0378}$ \\
 \hline
 ML(3) + UF &  0.043(4)  & 12.1  & 13.5 & {\bf 15.4} & {\bf 17.8} \\
 Lazy + UF & 0.031(3) & {\bf 11.1} & {\bf 12.8} & {16.6} & --- \\
UF & 0.037(8) & {11.5}  & {13.4}  & 15.7 & 18.9\\
 \hline 
 ML(3) + MWPM$^*$ &  {\bf 0.044(5)}  & 14.6 & {25.8} &  {81.5} &  {229} \\ 
 MWPM & 0.043(7) & 211  & 239& 273& 294\\
 \hline  
\end{tabular}

\caption{{\bf Overview of results.}
		For a number of variants of our decoding algorithm we provide the error threshold $p_{\rm th}$ (2nd column)
		for depolarizing noise (upper panel) and additional syndrome measurement errors (lower panel) 
		where ancillary qubits for measuring syndromes are also subject to depolarizing noise,
		as well as wall-clock time measurements (in milliseconds) of the decoding time for different error rates 
		(averaged over $10^6$ instances) for code distances $L=255$ and $L=31$, respectively.
		The bold-faced entries identify the best performing algorithm when optimizing for error threshold or compute times.
		Comparisons are shown for the union-find (UF) and minimum weight perfect matching (MWPM) decoders, combined 
		with either lazy \cite{delfosse2020hardware} 
		or machine learning (ML) assisted preprocessing using subsystems of size $\ell=3,5$ or $7$ as indicated in brackets (see main text).
		We have used a custom implementation for the UF decoder~\cite{UF_CYP} and PyMatching~\cite{PyMatching} for the MWPM.
		In the presence of additional syndrome errors, the pure MWPM calculation was optimized 
		by combining the Blossom and Dykstra algorithms  
		and for the ML-assisted MWPM with precomputed shortest paths.
		Details of our CPU/GPU hardware setup are provided in the appendix Sec. \ref{App:Benchmarking}. 
		\label{Tab:Overview}
}		
\end{table}

It is the purpose of this paper to introduce a powerful two-step decoding algorithm that combines neural network based preprocessing and union-find decoding to simultaneously achieve (i) improved error thresholds for depolarizing noise (even in the presence of syndrome measurement errors), (ii) algorithmic scalability up to tens of thousands of qubits, 
and (iii) real-life wall-clock run times (i.e.\ the elapsed time passed to execute the decoding process) that, for a range of error rates, best even those of the bare union-find algorithm, as summarized in Table \ref{Tab:Overview}. 
Our main algorithmic idea can be described as a hierarchical approach \cite{delfosse2020hardware} that employs an ML decoder to preprocess {\em local} error corrections and leave the unresolved longer-range errors to a conventional UF decoding. The preprocessing step shifts a two-dimensional subsystem over a given stabilizer code (akin to the preprocessing in a convolutional neural network often employed in image processing) and decodes local errors in these subsystems. After this step, the system still exhibits errors that require longer range corrections, for which we employ a conventional UF decoder. However, since the preprocessing reduces the effective error rate -- up to two orders of magnitude depending on the original error rate -- this second step is extremely performant as compared to, e.g., employing UF decoding to the original unprocessed error instances. 
Extensive wall-clock time measurements of our approach (the true performance indicator in many real-life applications) show that our algorithm
outperforms the bare UF decoder in a noise regime from 1\% (in which one might want to operate quantum computing devices) up to the 10\%  regime where our ML-assisted approach is found to push the error threshold by some 15 percent above the value of the bare UF decoder. 
Our approach bears some similarity to the `lazy UF decoder' \cite{delfosse2020hardware}, which also employs hierarchical decoding with a strictly local, hard decision preprocessing step and has been shown to substantially improve UF decoding for ultralow error rates below the per mil range.


\paragraph*{Hierarchical QEC.--}
Throughout the paper, we apply our decoding algorithm to the toric code in the presence of depolarizing noise
as well as a scenario with additional syndrome measurement errors.
For the latter, we use a phenomenological noise model where ancilla qubits for measuring syndromes are also subject to depolarizing noise but propagation of errors between data and ancilla qubits is neglected.
We consider a standard setup, where the toric code is defined on a square lattice of size $L \times L$ 
and the stabilizer operators around the vertices and plaquettes are given by $X_v = \prod_{i \in v} X_i$ and $Z_p = \prod_{i \in p} Z_i$. 
The code space is then spanned by the basis vectors $\{\ket{\psi}: X_v \ket{\psi} = 1 \: \forall v , Z_p \ket{\psi} = 1 \: \forall p \}$,
which, for periodic boundary conditions, is $4$ dimensional (and thus encodes $2$ qubits) 
and the distance of the code (i.e.\ the minimum length of a Pauli string that transforms one states in the code space to another state) is $L$. 
Each $Z$ ($X$) error on a qubit (located on the edges of the lattice) flips the value of the nearby $X_v$ ($Z_p$) operators.

\begin{figure}[t]
	\includegraphics[width=\columnwidth]{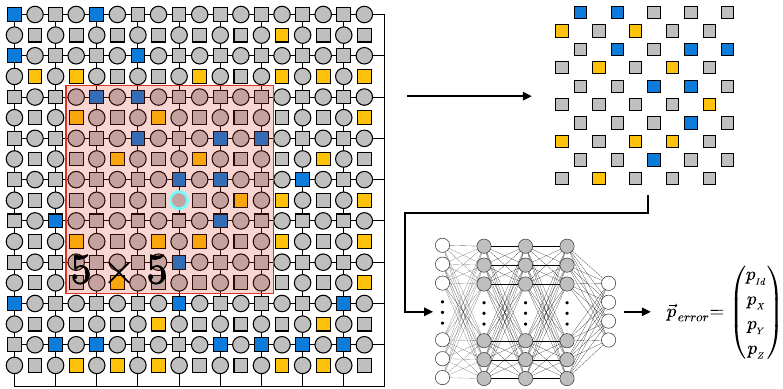}
	\caption{\textbf{Construction of the input data and neural network}. A single qubit (cyan circle) is selected as the reference point of the construction. All syndromes in the immediate vicinity (red square) are used as the input, whereby measured syndromes (blue/yellow) are assigned the value +1/-1 and not measured syndromes (grey) are assigned value 0, respectively. Passing the input through the feed forward network results in the error probabilities of the initially selected qubit.}
	\label{Fig:Setup}
\end{figure}

The decoding problem is then defined as identifying the error configuration for a given syndrome, 
i.e.\ a given measurement of the outcomes of all stabilizers $X_v$ and $Z_p$.
To do so, we employ a two-step hierarchical procedure. 
In the first stage -- the ML-assisted preprocessing --  
we aim to remove those errors that can be inferred from {\em local} syndromes. 
To this end, we only consider qubits directly connected to so-called defects 
(identified by an odd syndrome measurement $X_v = -1$ or $Z_p = -1$), 
as they are the typical source of locally correctable errors.
To infer which error is the most probable for a given qubit, 
our preprocessing step shifts through all qubits with a subsystem of size $\ell \times \ell$
 centered around an `examination qubit' located at its center (see the setup in Fig.~\ref{Fig:Setup}).
The local inference task for each such examination qubit is then assigned to a neural network, whose details we discuss below. 
The results of the inference are collected and the resulting corrections are applied in one shot at the end.
The result of this step is that a large number of local errors are decoded and only a small fraction of non-local errors, manifest on scales
beyond the range of our subsystem, remain.

The second stage of our algorithm is to then process these remaining non-local errors, which are left after updating the syndromes 
using the local error decoding in the previous stage. To do so, we employ a conventional UF decoder on the remaining syndrome. 
Doing so is significantly more efficient than employing the UF decoder on the bare decoding problem (without the preprocessing),
as we will see that the effective error rate for this UF decoding step is up to two orders of magnitude smaller than the original error rate 
(see the results section below).


\paragraph*{Neural decoder.--}
At the heart of our hierarchical QEC approach is a neural network that decodes error syndromes within a local subsystem, 
as illustrated in Fig.~\ref{Fig:Setup}.
We train this neural network to output the most probable error (among the four possible $\{I,X,Y,Z\}$ errors) of the central qubit  
given $2\ell^2$ nearby syndromes as an input (with the factor of 2 coming from the two types of $X$ and $Z$ measurements).
In machine learning, this type of task is commonly known as multiclass classification problem and exceedingly well-studied 
in the context of supervised learning approaches (e.g.\ in image classification).
To adopt such a supervised learning approach to optimize our neural network, we do training with a labeled dataset, 
i.e.\ batches of error-syndrome pairs generated for a given error rate (and noise model), training separate networks for each error rates. 
In doing so, we use the error of each qubit directly touching a defect (as in the inference step outlined above) as a label 
and $X$, $Z$ syndromes of size $\ell\times \ell$ surrounding this qubit as input.
In practice, we train our networks in $10^6$ epochs, for which we create independent sets of 512 error-syndrome batches `on the fly',
which also reduces the chance of overfitting. 

In designing the neural network architecture, we realize that there is an inherent trade-off between the two algorithmic layers
of our hierarchical approach: If one opts for a small neural network, its computation time remains low but its accuracy in resolving local syndromes drops, resulting in more computational load for the UF decoder on the higher algorithmic layer. If, on the other hand, one opts for a large neural network, its accuracy in resolving syndromes goes up at the cost of larger compute times, while also alleviating the load of the higher-level UF decoder. Indeed, this trade off leads to a sweet spot, i.e. an intermediate neural network size that results, e.g., in minimal wall-clock run times {\em or} maximal error thresholds. To identify an optimal configuration, we have explored a multitude of different network architectures for the case of depolarizing noise, varying the size of the subsystem, the depth of the network, and the number of nodes per layer as main parameters (as detailed in the appendix Sec. \ref{Apx:NeuralNetwork} ).
When optimizing for compute speed a $5 \times 5$ subsystem turns out to be ideal, while pushing the error threshold one might want to go with a $7 \times 7$ subsystem -- see Table \ref{Tab:Overview}. However, since the error threshold of the speed-optimized network is only 3\% smaller than the threshold-optimized network, we consider the $5 \times 5$ neural network approach the best compromise in achieving fast decoding {\em and} high error thresholds for an algorithm that {\em also} delivers on high scalability.


\begin{figure}[t]
	\includegraphics[width=0.99\columnwidth]{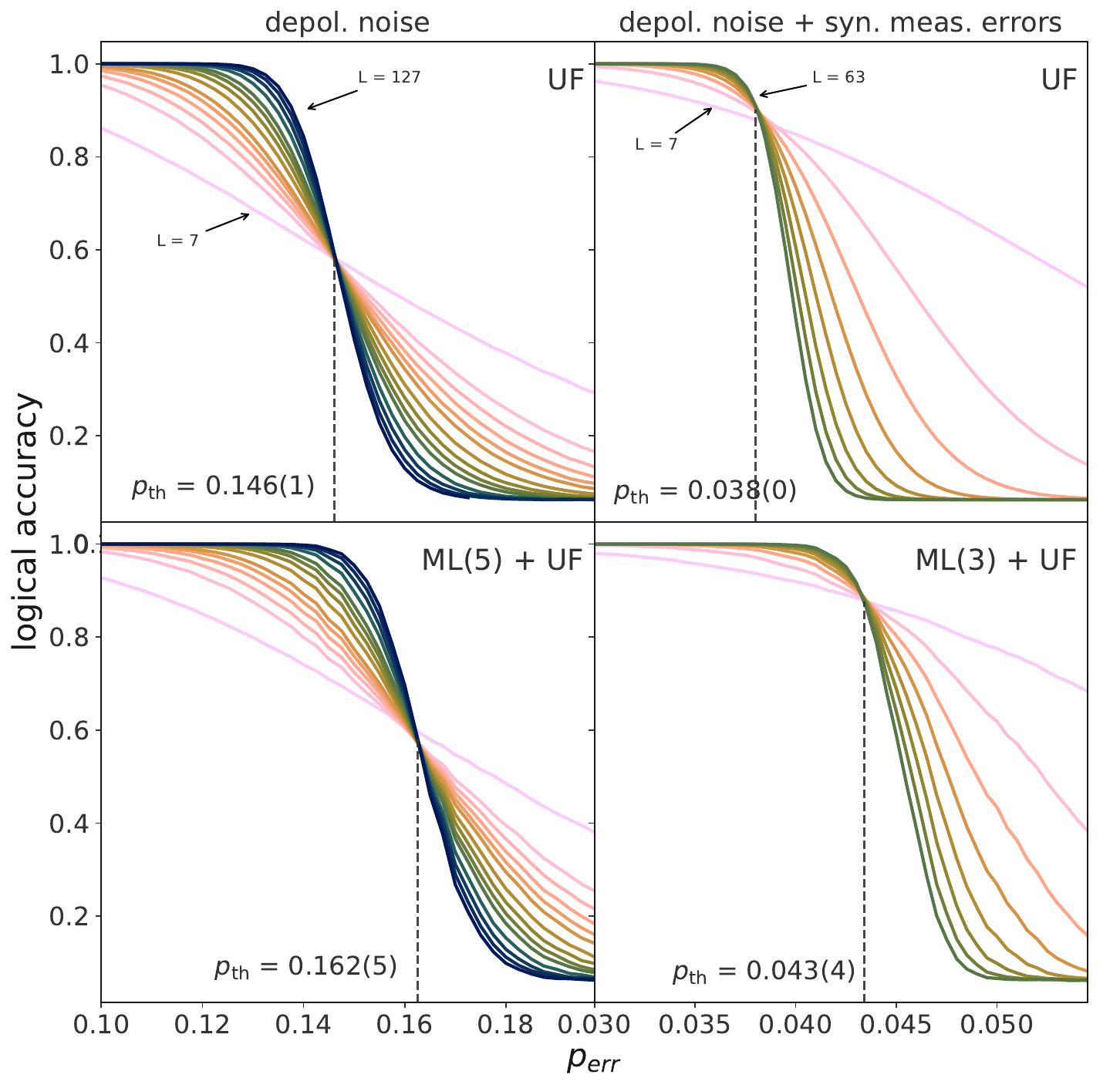}
	\caption{\textbf{Error threshold and scaling behavior} for the conventional union find (UF) algorithm (upper panel),
	and the machine learning assisted union find (ML+UF) algorithm (lower panel) for depolarizing noise. 
	The ML assisted algorithm shifts the error threshold by some 10\%, 
	from $p_{\rm err} = 0.146(1)$ for the UF to $p_{\rm err} = 0.162(5)$ for the ML+UF algorithm.
	}
	\label{Fig:Thresholds}
\end{figure}

\begin{figure}[t]
	\includegraphics[width=0.99\columnwidth]{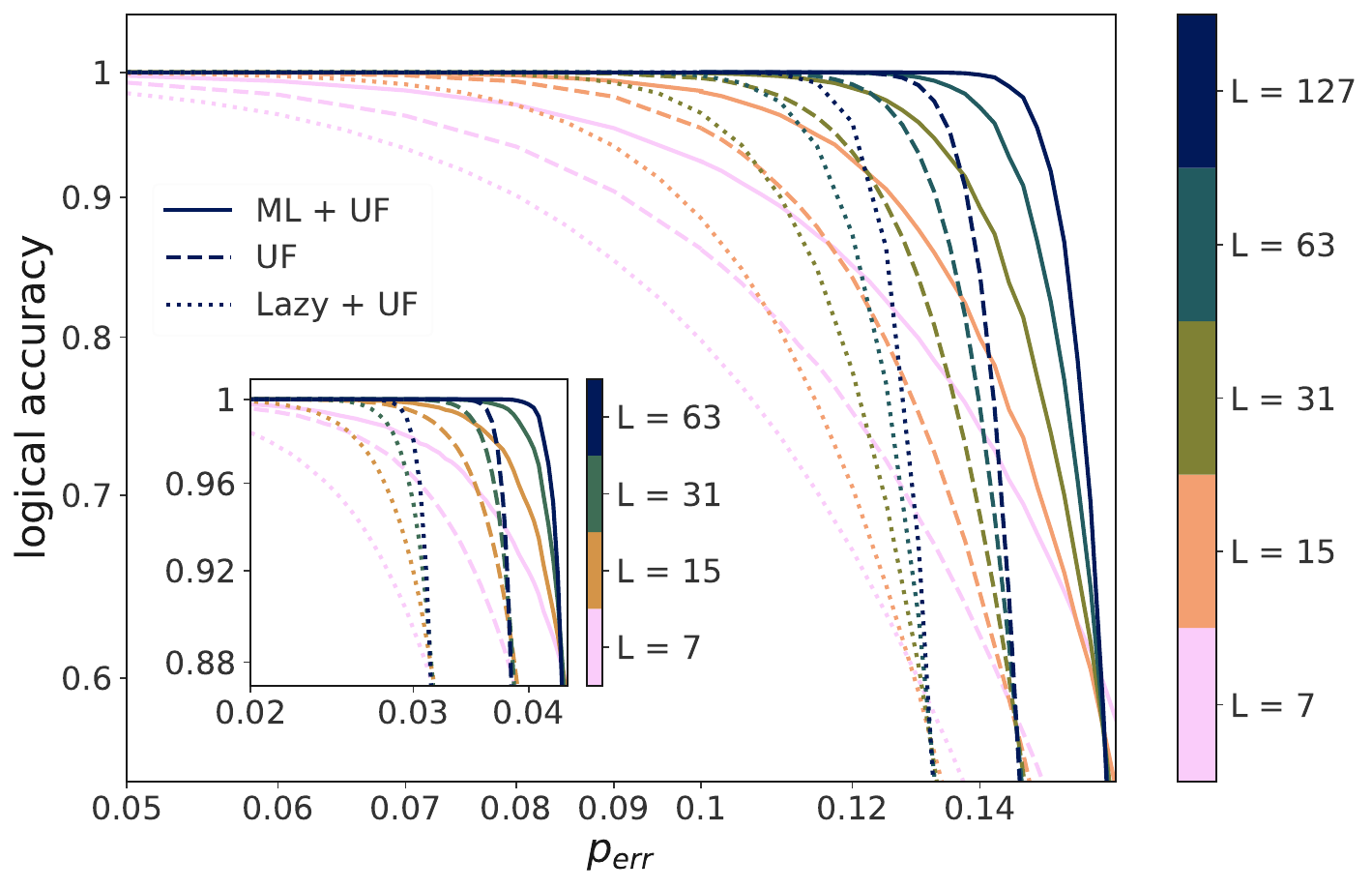}
	\caption{\textbf{Logical accuracy} of the conventional union find (UF) decoder and combined with lazy or ML-assisted preprocessing
	for depolarizing noise. The inset shows the case of additional syndrome measurement errors. 
	The ML+UF decoder increases the logical accuracy, independent of system size, for all error rates shown.
	}
	\label{Fig:LogicalAccuracy}
\end{figure}

\paragraph*{Benchmark results.--}
In benchmarking our hierarchical QEC algorithm, we start in the high-noise regime and calculate the error threshold of our approach. Decoding $10^6$ random instances of depolarizing noise for different error rates and linear system sizes in the range $L = 7, \ldots, 127$ 
we can readily deduce the error threshold from the finite-size scaling shown in Fig.~\ref{Fig:Thresholds}. 
In comparison to the bare UF algorithm (top panel), which exhibits an error threshold of $p^{\rm UF}_{\rm th} = 0.146(1)$,
our algorithm yields a 10\% higher value of $0.162(5)$ (when we employ a $5\times 5$ subsystem) and an increase of more than 20\% compared to the lazy UF decoder's threshold of $0.131(9)$
\footnote{
We found that one can slightly push the threshold value even further up (at the expense of additional compute time) 
by employing a larger $7\times 7$ subsystem, which gives $p_{\rm th} = 0.167(0)$ 
(see also Table \ref{Tab:Overview}). Going to even larger subsystems has not resulted in any further notable improvement of the threshold value
in our measurements.
}.
This notable increase of the error threshold indicates that our ML-assisted approach is capable of identifying and resolving
{\em correlated} errors in the depolarizing noise, which the bare UF decoder cannot handle. 
The strength of the ML-assisted decoder 
in the dense error regime can also be exemplified by the logical accuracy near the threshold plotted in Fig.~\ref{Fig:LogicalAccuracy}, 
which shows a higher logical accuracy for the ML+UF decoder in this regime, independent of system size.
It should further be noted that our threshold values are higher than the one of the bare RG decoder \cite{RG1} with $p^{\rm RG}_{\rm th}=0.153$ 
and comparable to those found for a combination of RG and sparse decoders \cite{Poulin2014}, 
or the best ML-based decoders using deep neural networks, 
for which  error thresholds of $p^{\rm ML}_{\rm th} \approx 0.165$ are reported \cite{Krastanov2017,Fitzek2020Deep} for depolarizing noise.
However, our result is still significantly below the optimal value of $p_{\rm opt} = 0.189(3)$, inferred from a mapping  \cite{Bombin2012} of the decoding problem to the classical disordered eight-vertex Ising model.
Performing a similar analysis for the scenario of depolarizing noise and syndrome measurement errors, we come to analogous conclusions with a spread of the error threshold between $p_{\rm th} = 0.031(3)$ for the lazy UF decoder and 0.044(5) obtained for ML-assisted MWPM decoding (lower panel of Table \ref{Tab:Overview}).

\begin{figure}[b]
	\centering
	\includegraphics[width=0.99\linewidth]{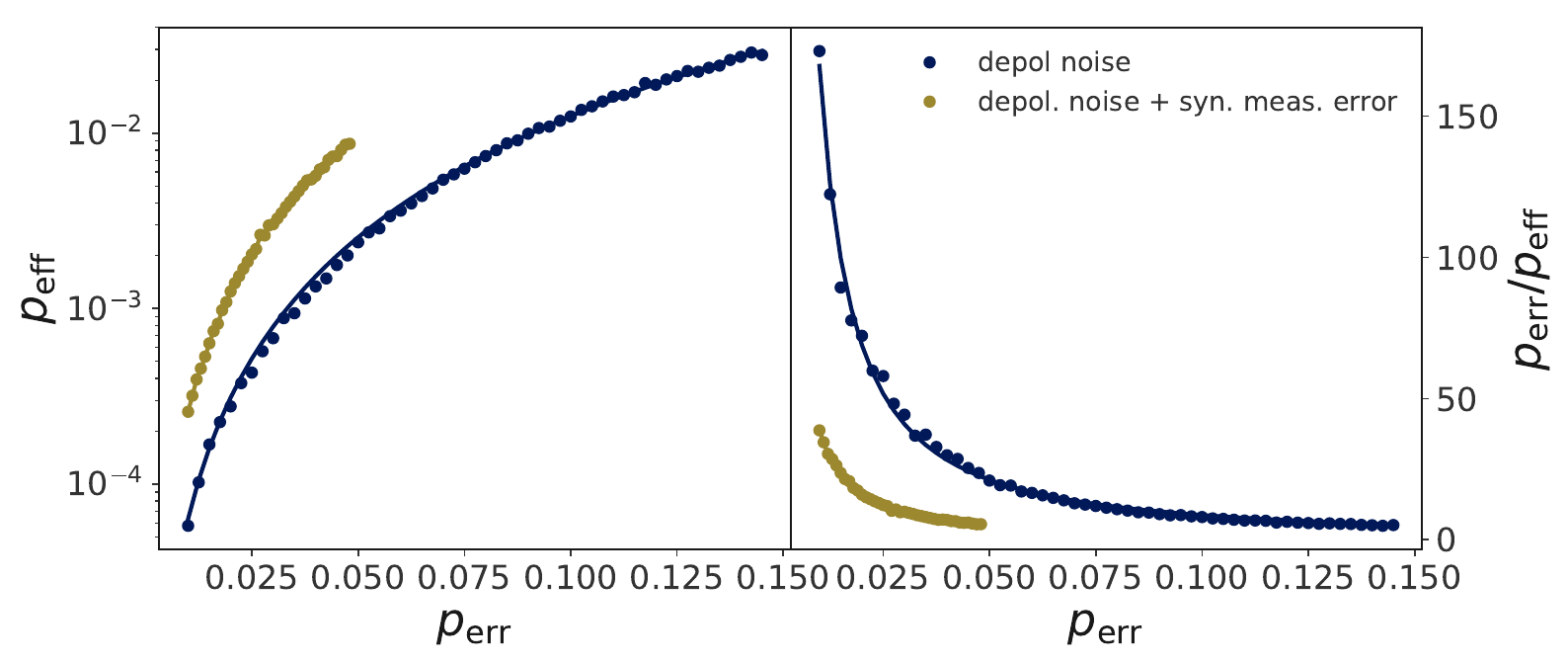}
	\caption{\textbf{Effective error reduction} attained by the ML preprocessing step of our hierarchical algorithm.
	The left panel shows the effective error rate $p_{\rm eff}$ as a function of the original error probability $p_{\rm err}$,
	illustrating the effect of local decodings in the preprocessing step. 
	The effective error rate is calculated 
	from the number of remaining syndromes 
	$p_{\rm eff} = \sum S_i / ( \frac{4}{3} \times 2 \times 2L^2)$.
	The right panel shows the ratio of the original error probability and the effective error rate.
	}
	\label{Fig:EffectiveErrorRate}
\end{figure}

One measure to illustrate the inner workings of our hierarchical approach is an `effective error rate', i.e.\ the 
reduction of errors obtained after performing the first step of our algorithm -- the preprocessing using a subsystem and neural network-based local decoding.
Shown in Fig.~\ref{Fig:EffectiveErrorRate}, this effective error rate reveals that this preprocessing step is particularly powerful at low error rates,
i.e. in the regime where few long-range errors occur. 
Here one can reduce the initial error rate by more than two orders of magnitude (see the right panel of Fig.~\ref{Fig:EffectiveErrorRate}), 
thereby significantly speeding up the subsequent UF decoding step (as compared to a direct application to the original syndrome).

\begin{figure}[t]
	\centering
	\includegraphics[width=0.99\linewidth]{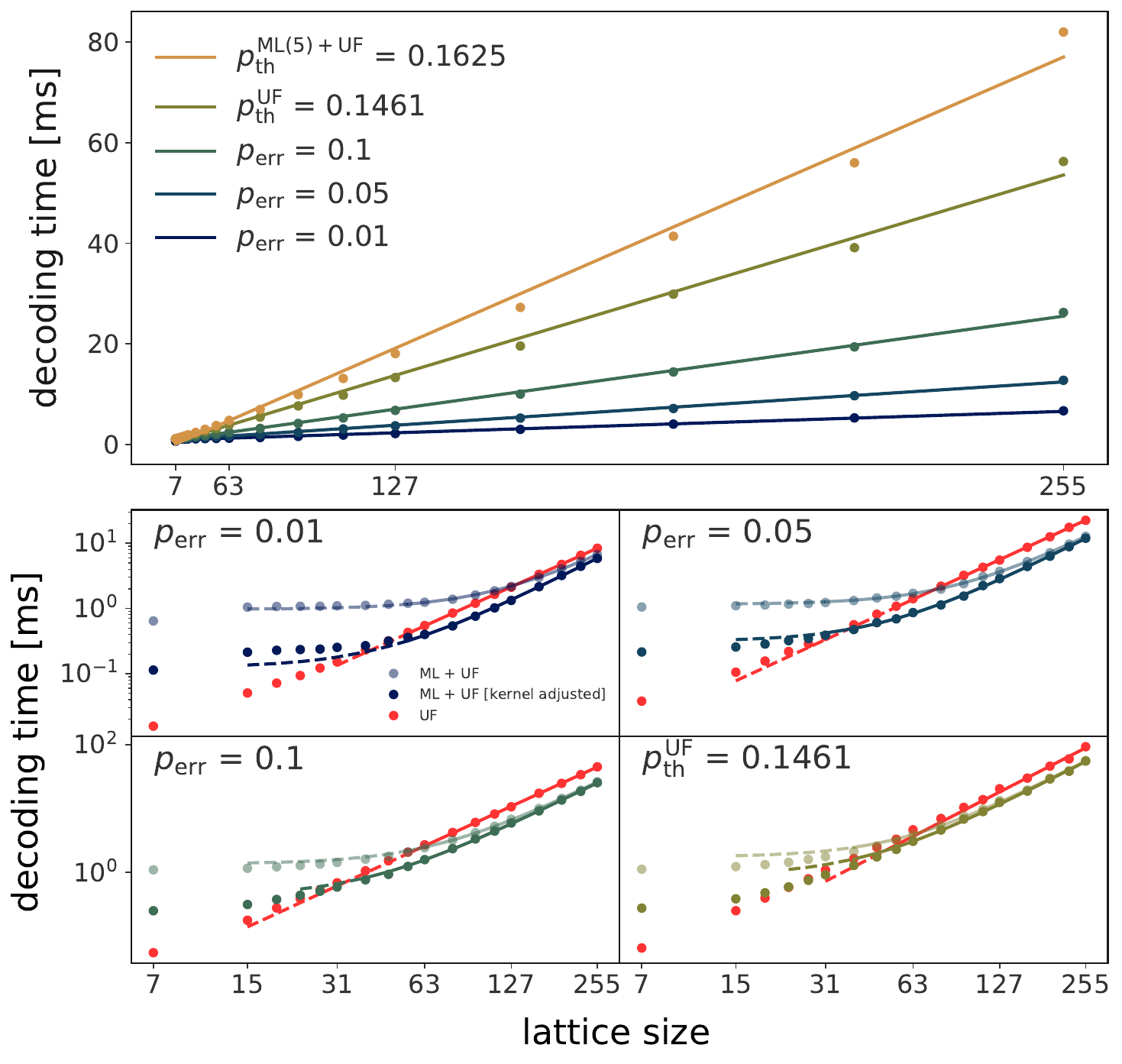}
	\caption{
		\textbf{Algorithmic scaling} of our hierarchical decoder for various  rates of depolarizing noise (top panel).
		The bottom panels show detailed comparisons with the bare UF decoder on a log-log scale.
		Shown in all panels is the average decoding time measured in wall clock time averaged over $10^6$ error instances.
		The `kernel adjusted' time in the lower panels is the ML+UF decoding time subtracted by a constant offset, 
		to compensate kernel launch times (see main text).
		 }
	\label{Fig:Scaling}
\end{figure}

As such one might naively expect the biggest computing gain of our algorithm in the low-noise regime. 
For practical implementations this is, however, not true as becomes apparent when performing run time measurements of our decoder. 
Such measurements are illustrated in Fig.~\ref{Fig:Scaling} where the decoding time (again averaged over $10^6$ error instances) is plotted versus the linear system size for different error rates. 
The top panel nicely demonstrates that, for large system sizes, we find near linear scaling for both the UF and our hierarchical UF+ML decoder, independent of the error rate. Note that our ML-assisted decoder easily scales up to $2 \times 255 \times 255 \approx 130,000$ qubits where 
the decoding time per instance is still a fraction of a second -- this should be contrasted to other ML-based decoders reported in the literature,
which could not be scaled beyond a hundred qubits (see the overview in the appendix Table \ref{Tab:Training}). 

If we look at the scaling of our algorithm for small to moderate system sizes (highlighted in the lower panels of Fig.~\ref{Fig:Scaling}), 
a breakdown of the linear scaling of the ML-assisted decoder becomes evident. There is indeed a considerable `lag' in the implementation of our hierarchical approach, which arises from the usage of an external GPU to perform the neural network-based preprocessing step (see appendix Sec. \ref{App:Benchmarking} for hardware specifications). Doing so readily implies another inherent trade off: Initializing the neural network and loading the syndrome data to the GPU component has an almost constant overhead, which explains the plateau in our scaling plots for small system sizes where the advantage of fast GPU processing of the neural network is not compensating this overhead (as it is the case for large system sizes). We have measured this `kernel start-up' time as to subtract parts of this overhead -- which would not exist in a dedicated or in-situ device in a practical implementation of QEC in the lab -- to arrive at the `kernel adjusted' scaling curve in Fig.~\ref{Fig:Scaling}. The crossing point at which the ML-assisted decoder outperforms the bare UF decoder in this setup comes down to code distances of $L\approx 31$, but we expect even smaller code distances to benefit from the ML-assisted approach when going for an in-situ implementation \cite{das2020scalable}, e.g. using FPGAs or TPUs \cite{Jouppi2017TPU}.

In summary, we have demonstrated that the combination of machine-learning assisted preprocessing in conjunction with conventional decoders 
in a newly devised hierarchical approach results in a vastly scalable algorithm. Our practical implementation shows that one can increase logical accuracy and push the error threshold by resolving correlated errors, while also reducing the actual decoding times (to a few milliseconds in our hardware setup) particularly in the dense error regime. 
As such our approach nicely complements the lazy UF decoder  \cite{delfosse2020hardware} which excels in the opposite regime of ultralow error rates. Taken together, one might argue that one should always combine the UF decoder with some sort of preprocessing step -- which one to go for depends on the expected noise level and code distances.

\paragraph*{Acknowledgments.--}
We thank M. Kastoryano and T. Wagner for insightful discussions, as well as O. Higgott for comments on optimizing the PyMatching
results for MWPM decoding the phenomenological noise model.  
This  project  was  funded  by  the  Deutsche  Forschungsgemeinschaft 
under Germany's Excellence Strategy - Cluster of  Excellence  Matter  and  Light  for  Quantum  Computing  (ML4Q)  EXC  2004/1  - 390534769
and within the CRC network TR 183 (project grant 277101999) as part of project B01.\\

\paragraph*{Note added.--}
In the days prior to submission of this manuscript a tensor network decoder has been introduced \cite{chubb2021general},
for which  an error threshold for depolarizing noise in the toric code of $p_{\rm th} = 0.1881(3)$ is reported, indistinguishable 
within error bars from the upper bound inferred from a mapping of the decoding problem to the classical disordered eight-vertex 
Ising model \cite{Bombin2012}.

\nocite{TensorFlow}
\nocite{Gabos1974Implementation,Lawler1976Combinatorial}
\nocite{Fowler2012Towards,Higgott2020Subsystem}
\nocite{Delfosse2020Linear}


\bibliography{bibliography}

\appendix


\section{Neural Networks}
\label{Apx:NeuralNetwork}

This appendix provides a detailed overview of the tested network architectures to find an optimal decoding setup,
along with an expose of the  hyperparameters used in the training of the neural network(s). 

\subsection*{Network Architecture}

\begin{table}[b]
\begin{tabular}{ |c|c|c|c|c| }
\hline
 hidden  & hidden  & \multicolumn{3}{c|}{total parameters} \\
  layers  &  nodes  &  $\ell = 3 $ &  $\ell = 5 $ &  $\ell = 7 $  \\
 \hline
 1 & 32 	& 740 	&  1764 	&  3300\\
 1 & 64 	& 1476 	&  3526 	&  6596\\
 1 & 128 	& 2948 	&  7044 	& 13188\\
 1 & 256 	& 5892 	& 14084 	&  26372\\
 \hline
 2 & 32 	& 1796 	& 2820 	& 4356\\
 2 & 64 	& 5636 	& 7684 	& 10756\\
 2 & 128 	& 19460 	& 23556 	& 29700\\
 2 & 256 	& 71684 	& 79876 	& 92164\\
 \hline
 3 & 32 	& 2852 	& 3876	& 5412\\
 3 & 64 	& 9796 	& 11844 	& 14916\\
 \textbf{3} & \textbf{128} 	& 35972 	& \textbf{40068} 	& 46212\\
 3 & 256 	& 137476 	& 145668 	& 157956\\
 \hline
\end{tabular}
\caption{{\bf Overview of neural network architectures} tested to solely optimize the wall-clock run time by varying the size of the 
$\ell \times  \ell$ subsystem  (see Fig. 1 of the main text), the number of hidden layers, and nodes per layer.
The optimal configuration is highlighted in bold-face. 
Note that when vying to solely maximizing the error threshold, a different network architecture is singled out as ideal (see main text).
\label{Tab:NetworkArchitectures}
}
\end{table}

In order to optimize our hierarchical algorithm to simultaneously achieve minimal wall-clock run times and high error thresholds, while remaining scalable, we have explored a multitude of different network architectures, varying the size of the subsystem, the depth of the network, and the number of nodes per layer as main parameters. A systematic overview of some of the tested network architectures is provided in Table~\ref{Tab:NetworkArchitectures} below.

As discussed in the main text, our hierarchical approach has an inherent trade off between the two hierarchical algorithmic layers: 
If one opts for a small neural network, its computation time remains low but its accuracy in resolving syndromes drops, resulting in more computational load for the UF decoder on the higher algorithmic layer. If, on the other hand, one opts for a large neural network, its accuracy in resolving syndromes goes up at the cost of larger compute times, while also alleviating the load of the higher-level UF decoder. 
Indeed, this trade off leads to a sweet spot, i.e. an intermediate neural network size that results, e.g., in optimal compute time.

\begin{table}[t]
\begin{tabular}{ |c||c|  }
 \hline 
 \multicolumn{2}{|c|}{depolarizing noise} \\
 \hline 
 \hline 
 \multicolumn{2}{|c|}{$\ell = 5$ network parameters (optimal wall-clock time)} \\
 \hline
 hidden layers & 3\\
 hidden nodes per layer & 128\\
 total number free parameter & 40 068\\
 activation functions  hidden layer & Relu \\
 activation functions  output layer & Softmax \\
 \hline
 \multicolumn{2}{|c|}{$\ell = 7$ network parameters (optimal error threshold)} \\
 \hline
 hidden layers & 5\\
 hidden nodes per layer & 512\\
 total number free parameter & 1 103 364\\
 activation functions  hidden layer & Relu \\
 activation functions  output layer & Softmax \\
 \hline
\end{tabular}
 
 \vskip 3mm
 
\begin{tabular}{ |c||c|  }
 \hline
\multicolumn{2}{|c|}{depolarizing noise + syndrome measurement errors} \\
 \hline
 \hline 
  \multicolumn{2}{|c|}{$\ell = 3$ network parameters (optimal wall-clock time)} \\
 \hline
 hidden layers & 3\\
 hidden nodes per layer & 128\\
 total number free parameter & 43 396\\
 activation functions  hidden layer & Relu \\
 activation functions  output layer & Softmax \\
 \hline

\end{tabular}
\caption{{\bf Optimized network architectures.} 
		The upper two panels provide, for depolarizing noise, 
		detailed parameters for a speed-optimized decoder (minimizing the wall-clock time), 
		based on a $5 \times 5$ subsystem neural network 
		and a threshold-optimized decoder, based on a $7\times 7$ subsystem neural network.
		The lower panel, for depolarizing noise and syndrome measurement errors, 
		provides detailed parameters for a 3-dimensional adaptation of the speed-optimized decoder 
		in this effectively three-dimensional settings with a $3 \times 3 \times 3$ octahedral subsystem  
		as illustrated in Fig. \ref{Fig:3DSetup}.
	 \label{Tab:OptimalNetworks}
}
\end{table}

If our goal is to optimize for {\sl decoding speed}, i.e. minimizing the wall-clock time measured for decoding $10^6$ random instances for a toric code with linear system size $L=255$ (corresponding to the values listed in Table I of the main manuscript), we find that among the multitude of different network architectures listed in Table~\ref{Tab:NetworkArchitectures}, the $5 \times 5$ neural network with 3 hidden layers and 128 nodes per layer (resulting in a total of 40068 adjustable parameters) is the best choice (see also Table \ref{Tab:Training}).

If, on the other hand, our goal is to optimize the {\sl decoding efficiency}, i.e.maximizing the error threshold, then we find that a $7 \times 7$ neural network with 5 hidden layers and 512 nodes per layer (resulting in more than $10^6$ adjustable parameters in total) is the most favorable network configuration. 

However, we note that the error threshold of the speed-optimized network already comes in at $p_{\rm th} = 0.162(5)$ and as such is only slightly smaller than the maximal threshold of $p_{\rm th} = 0.167(0)$, which we find for the efficiency-optimized network -- in particular when comparing this to the threshold values of the conventional decoders (with error thresholds around $p_{\rm th} \approx 0.150 \pm 0.004$) and the ideal threshold  $p_{\rm th} = 0.189$ inferred from a mapping of the decoding problem to the classical disordered eight-vertex Ising model \cite{Bombin2012}. As such, we conclude that the speed-optimized $5 \times 5$ neural network already presents a good compromise in achieving fast decoding {\sl and} high error thresholds for an algorithm that also delivers on high scalability.

\begin{figure}[t]
	\includegraphics[width=0.99\columnwidth]{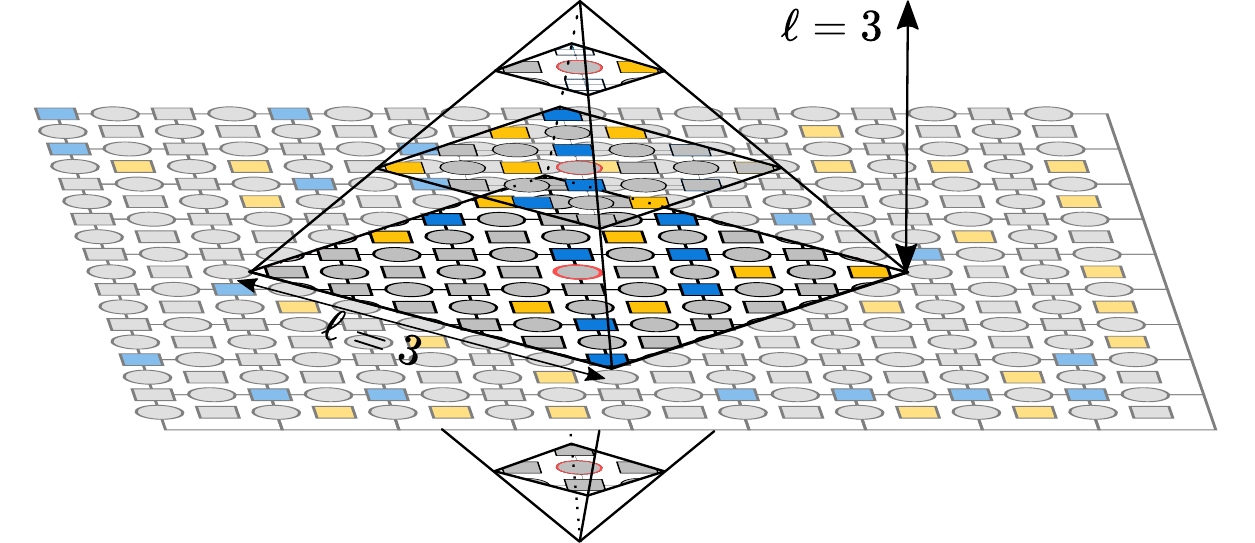}
	\caption{\textbf{Three-dimensional subsystem.} 
		For the augmented noise model where we pair depolarizing noise with syndrome measurement errors
		we apply an octahedral subsystem in lieu of the two-dimensional setting described in Fig.~1 of the main text.
		}
	\label{Fig:3DSetup}
\end{figure}

\subsection*{Training}

Each neural network configuration has been trained using standard supervised learning techniques, using labeled data sets of syndromes and their corresponding underlying errors For each error rate we train a separate neural network, using only error-syndrome pairs generated at that given rate. Because such error-syndromes pairs can be generated in very efficient ways, we have generated, for every epoch of the training, a new set of 512 such pairs `on the fly'. In effect, this procedure increases the size of the training dataset to be of the order `batch size $\times$ training epochs', and significantly reduces the chance of overfitting. Detailed training parameters are provided in Table~\ref{Tab:Training}, which have been optimized (via a grid search) for the error threshold probability of the decoder and then employed to every training process regardless of the underlying error probability.

\begin{table}[b]
\begin{tabular}{ |c||c|  }
 \hline
 \multicolumn{2}{|c|}{training parameters} \\
 \hline
 batch size   & 512\\
 epochs &   $10^6$\\
 learning rate& 0.001\\
 optimizer & ADAM\\
 loss function&   categorical cross-entropy\\
 training lattice size $L_{\rm train}$ & 7\\
 section distance $\ell$ & 5\\
 \hline 
\end{tabular}
\caption{{\bf Hyperparameters} used for the training of our various network architectures.
		 An epoch corresponds to the number of generated training batches (see text).
	 \label{Tab:Training}
}
\end{table}


\section{Benchmarking}
\label{App:Benchmarking}

In performing our wall-clock run time benchmarks to provide a measure for real-life applicability we have employed a hardware setup
based on the following CPU/GPU tandem for the depolarizing noise measurements
\begin{itemize}
  \item CPU: Intel Xeon CPU E5-2699A v4 @ 2.40~GHz,
  \item GPU: Nvidia Tesla V100 SXM2.
\end{itemize}
and for the measurements of the depolarizing noise plus syndrome measurement error model
\begin{itemize}
  \item CPU: AMD EPYC Rome 7402 CPU @ 2.80~GHz,
  \item GPU: Nvidia A100 Tensor Core-GPU.
\end{itemize}

The times measured are the bare times needed to decode a given syndrome and apply the error correction. 
As discussed in the main text, our GPU-assisted calculations for the machine learning parts include a `kernel time', 
i.e. the time needed to launch the CUDA and TensorFlow \cite{TensorFlow} kernel, every time a syndrome is decoded. 
We estimate this kernel time as the GPU time needed to run a neural network with subsystem $\ell=7$ 
for a relatively small error rate of $p_{\rm err}=0.01$, where the kernel launch times dominates over the data transfer time and actual 
GPU calculation time.


\section{Minimum Weight Perfect Matching (MWPM)}
\label{App:MWPM}

To make this manuscript self-contained, we also provide an error threshold calculation of the minimum weight perfect matching (MWPM) decoder
for the depolarized noise model using PyMatching~\cite{PyMatching} -- a particularly efficient implementation of MWPM decoder. 
The most widely used MWPM algorithm for general graphs is the one by Edmonds~\cite{MWPM}, which is also widely known as the Blossom algorithm. The time complexity of the original Blossom algorithm is $\mathcal{O}(|E||V|^2)$ where $|E|$ and $|V|$ are the number of edges and vertices, respectively.
In the toric code set-up, which we have considered throughout the paper, a simple implementation gives $O(n^4)$ (where $n=2L^2$ is the total number of qubits) as $|V| \propto n$ (number of qubits) and $|E| \propto n^2$. 
Fortunately, the algorithm for MWPM has been improved and some suggest $\mathcal{O}(|V|^3)$~\cite{Gabos1974Implementation,Lawler1976Combinatorial} which gives $\mathcal{O}(n^3)$ for the toric code.
We thus consider this as the worst-time complexity of the decoding problem using the MWPM decoder.

\begin{figure}[t]
	\includegraphics[width=0.99\columnwidth]{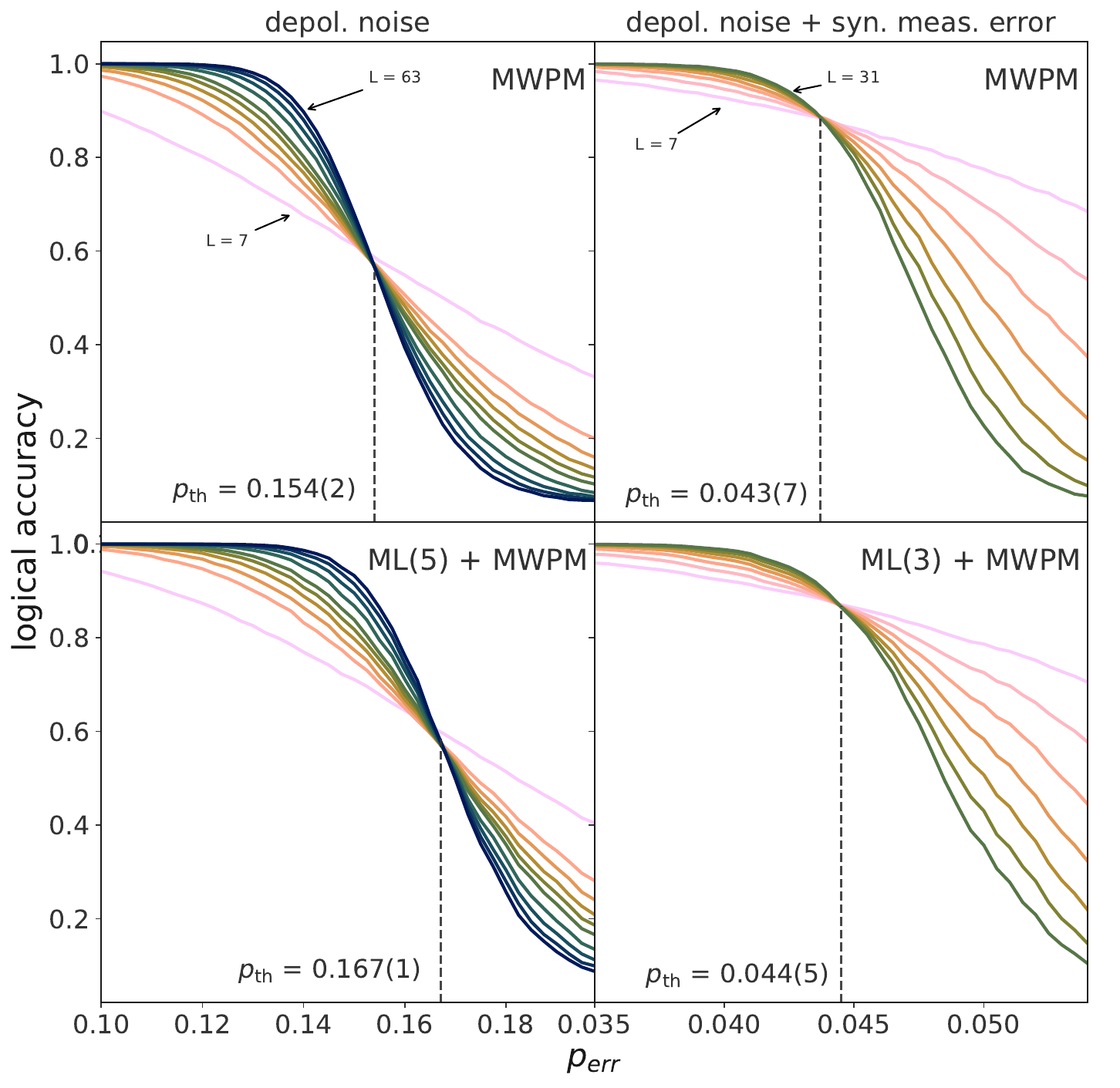}
	\caption{{Error threshold and scaling behavior} for the conventional {\bf minimum weight perfect matching (MWPM)} algorithm 
	(upper panel),
	and the {\bf machine learning assisted minimum weight perfect matching (ML+MWPM)} algorithm (lower panel) for depolarized noise (left)
	and an augmented noise model with additional syndrome measurement errors (right).
	The ML assisted algorithm shifts the error threshold by some 10\%, 
	from $p_{\rm err} = 0.1542$ for the bare MWPM decoder to $p_{\rm err} = 0.1671$ for the ML assisted MWPM algorithm (lower panel).}
	\label{Fig:ErrorThresholdMWPM}
\end{figure}

However, for decoding the toric code, an even more substantial speed up is possible by adopting the idea of locality -- instead of constructing edges between {\em all} defects, one may construct edges only between nearby defects within some constant distance. Such an idea has been considered in Refs.~\cite{Fowler2012Towards,Higgott2020Subsystem} and gave a visible speed-up for the decoding problem albeit this may potentially harm the threshold.
PyMatching~\cite{PyMatching}, which we have utilized for the MWPM decoder in this paper, could achieve $\mathcal{O}(L^{2.11})$ in benchmarks by combining this idea with an efficient C++ implementation of graph algorithms.

We present the threshold result for PyMatching in the upper panel of Fig.~\ref{Fig:ErrorThresholdMWPM}, where we find an error threshold of $p_{\rm err} = 0.1542 \pm 0.0006$.
This value is consistent with the previously established estimate of $3/2 \cdot p_{\rm err}^{\rm indep} \approx 0.1546$ (where $p_{\rm err}^{\rm indep} = 0.1031 \pm 0.0001$ is the threshold for the independent noise model estimated in Ref.~\cite{harrington2004analysis}).
This confirms that constructing edges only between nearby edges (as employed in PyMatching) barely shifts the threshold.

\begin{figure}[t]
	\includegraphics[width=0.99\columnwidth]{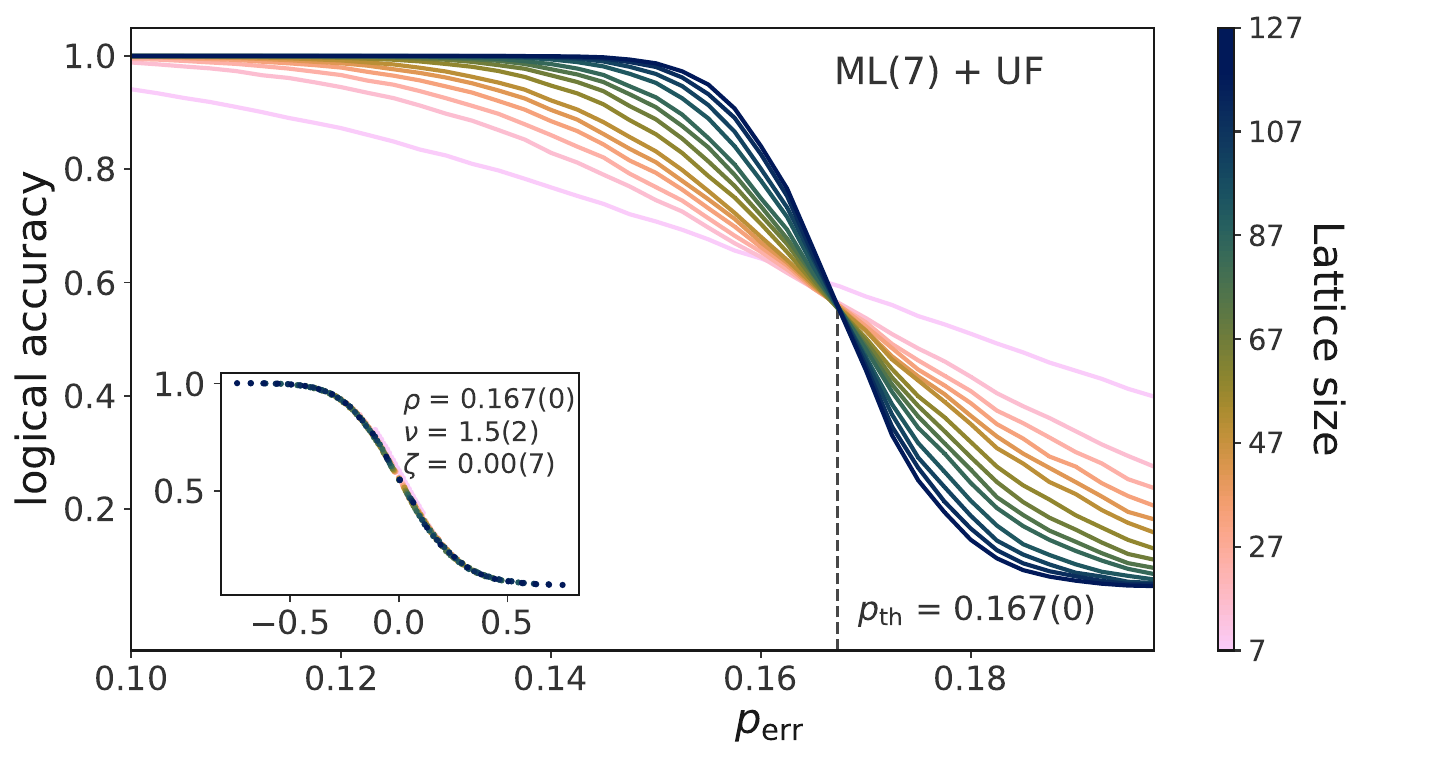}
	\caption{{Error threshold and scaling behavior} for the {\bf machine learning assisted union find (ML+UF)} algorithm 
	using a $7 \times 7$ subsystem. In comparison to the more time-efficient $5 \times 5$ subsystem of Fig. 2
	in the main text, the larger subsystem  applied here leads to a small shift of the error threshold to $p_{\rm err} = 0.167(0)$.
	The insets show a data collapse around this threshold value for different system sizes, consistently yielding a critical exponent
	of $\nu = 1.5(2)$. 
	}
	\label{Fig:Thresholds_large}
\end{figure}

We have also explored using the MWPM decoder within the hierarchical approach discussed in the main text (where the UF decoder is applied) in the lower panel of Fig.~\ref{Fig:ErrorThresholdMWPM}.
Similar to our results for the UF decoder,  we find a shift of the error threshold by some 10\% to a value of $p_{\rm err} = 0.1671 \pm 0.0005$. 
While these observations show that our hierarchical approach can be employed with different `conventional' decoders 
to achieve improved performance, we settled on the UF decoder for its algorithmic scalability bests all other conventional approaches.


\section{Union-find (UF) decoder} 
\label{App:UF}

\begin{figure}[t]
	\includegraphics[width=0.99\columnwidth]{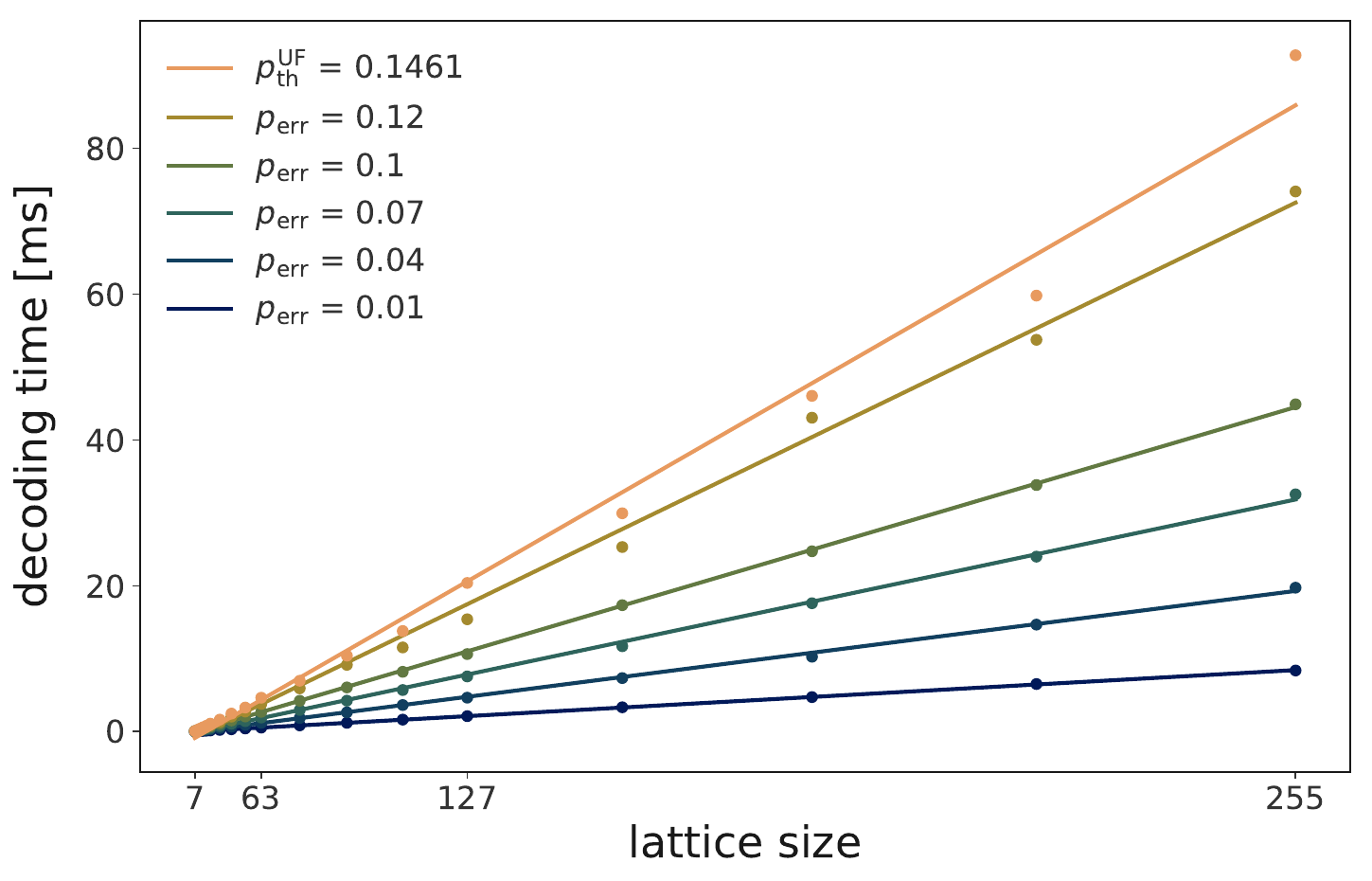}
	\caption{\textbf{Scaling of the UF decoder.} 
		Average decoding time of our custom implementation of the UF decoder~\cite{UF_CYP} as a function of linear lattice size. 
		We find a {\em linear} scaling for a wide range of error rates (the lines are linear fits to the data as a guide to the eye). 
		For the error rates near the threshold ($p_{\rm err} = 0.12$ and $p_{\rm th}$ in this plot), we find deviations from the linear fit,
		which we believe to originate from rehashing operations, 
		as our implementation relies on a {\sl hash set} data structure to maintain border vertices of a cluster.
		}
	\label{Fig:ScalingUF}
\end{figure}

Let us first briefly summarize the UF decoding algorithm \cite{UnionFind2017}.
For simplicity, we consider $Z$ errors so that each syndrome ($X_v$) is assigned to a vertex 
(when considering $X$ errors, the algorithm needs to be applied on the dual lattice).
The first step of the UF decoder is to iteratively construct clusters such that each cluster contains an even number of syndromes.
To do this, one first generates clusters where each one contains a single defect vertex. 
Then one pursues the following two steps iteratively until all clusters contain an even number of defects:
(i) Extend all clusters a half-edge in all four directions ({\sl Grow} step).
(ii) When two clusters meet at an edge, one merges the two clusters ({\sl Merge} step). 
When these iterations terminate, one can then use a peeling decoder~\cite{Delfosse2020Linear} 
to infer the error corrections from the constructed clusters.
To achieve an algorithmic scaling of $\mathcal{O}(n \alpha (n))$ (where $\alpha (n)$ is the inverse of Ackermann's function \cite{UnionFind2017}), 
one needs a disjoint-set data structure (which is also often called {\sl union-find}) to maintain data of all clusters, 
as well as a clever updating scheme for boundary vertices for each cluster. 
For more details on these algorithmic details, we refer to the original paper~\cite{UnionFind2017}.

\begin{table*}
  \begin{tabular}{||m{6.4cm} |M{1.2cm} |M{0.8cm} |M{1.6cm} |M{1.6cm}| M{1cm} |m{3.6cm}||}
  \hline
  \centering\arraybackslash title & date & QECC & noise model & $p_{\rm th}$ & $d_{\rm max}$ & \centering\arraybackslash algorithmic scaling \\ 
  \hline 
   \multicolumn{7}{c}{} \\
   \multicolumn{7}{c}{conventional decoders (non-ML)} \\
   \multicolumn{7}{c}{} \\
  \hline
  Analysis of quantum error-correcting codes: symplectic lattice codes and toric codes \cite{harrington2004analysis} {\bf (MWPM)} & 05/2004 & TC & DP$^*$& 0.154$^*$ & 63$^*$ & $\mathcal{O}(n^3)$ (worst-case) $\mathcal{O}(d^{2.11})$ \, (PyMatching~\cite{PyMatching})  \\
  \hline 
  Fast Decoders for Topological Quantum Codes \cite{RG1} {\bf (RG-decoder)} & 02/2010 & TC& DP& 0.164& 128 & $\mathcal{O}(d^2  \log d)$ \;\;\;\;\;\;\; (serial) $\mathcal{O}(\log d)$ \;\;\;\;\;\;\;\;\;\;\;\;\; (parallel) \\
  \hline 
  Almost-linear time decoding algorithm for topological codes \cite{UnionFind2017} {\bf (UF-decoder)} & 09/2017 & TC& DP$^*$& 0.146$^*$ & 127$^*$ & $\mathcal{O}(n \cdot \alpha(n))$  \\ 
  \hline
  General tensor network decoding of 2D Pauli codes \cite{chubb2021general} {\bf (TN-decoder)} & 01/2021 & TC& DP& 0.1881(3) & 64 & $\mathcal{O}(n  \log n+ n \chi^3)$  \\ 
  \hline
   \multicolumn{7}{c}{} \\
   \multicolumn{7}{c}{ML-assisted decoders} \\
   \multicolumn{7}{c}{} \\
  \hline
  Decoding Small Surface Codes with Feedforward Neural Networks \cite{Varsamopoulos_2017} & 05/2017 & SC & DP & $\sim $ 0.15 & 7 & $\mathcal{O}({\rm MWPM})$  \\
  \hline
  Neural Decoder for Topological Codes \cite{Torlai2017Neural} &  07/2017 & TC & i.i.d bit flip & $\sim$ 0.110 & 6 & $> \mathcal{O}({\rm MWPM})$  \\ 
  \hline
  Deep Neural Network Probabilistic Decoder for Stabilizer Codes \cite{Krastanov2017} & 09/2017 & TC & DP & 0.164 & 11 & $\gg \mathcal{O}({\rm MWPM})$\\
  \hline  
  Deep neural decoders for near term fault-tolerant experiments \cite{Chamberland_2018} & 07/2018 & SC & CLN & $\epsilon \sim 7.11 \times 10^{-4}$ & 5 & n.a. \\  
  \hline 
  Neural network decoder for topological color codes with circuit level noise \cite{Baireuther_2019} & 01/2019 & CC & CLN & $\epsilon \sim 0.0023$ & 7 & n.a. \\
  \hline  
  Neural Belief-Propagation Decoders for Quantum Error-Correcting Codes \cite{Liu2019Neural} & 05/2019 &TC& i.i.d. X\&Z & $\sim 0.07$ & 10 & n.a. \\
  \hline 
  Quantum error correction for the toric code using deep reinforcement learning \cite{Andreasson2019quantumerror} & 09/2019 &TC& i.i.d. bit flip& $\sim 0.1$ & 7 & $\gg \mathcal{O}(n)$ $\quad\quad\quad$\,\, (estimate) \\
  \hline
  Symmetries for a High Level Neural Decoder on the Toric Code \cite{Wagner2020Symmetries} & 10/2019 &TC& DP & n.a. & 7 &  $\mathcal{O}({\rm MWPM})$ \\
  \hline
  Deep Q-learning decoder for depolarizing noise on the toric code \cite{Fitzek2020Deep} & 05/2020 &TC& DP & $\sim 0.165$ & 7 (9) & $\gg \mathcal{O}(n)$ $\quad\quad\quad$\,\, (estimate) \\
  \hline
  Reinforcement learning for optimal error correction of toric codes \cite{Domingo2020Reinforcement} &  06/2020 & TC & i.i.d bit flip &0.103 & 9 & $> \mathcal{O}({\rm MWPM})$  \\ 
  \hline
  Neural Network Decoders for Large-Distance 2D Toric Codes \cite{Ni2020neuralnetwork}& 08/2020 &TC& i.i.d. bit flip& $\sim 0.103$ & 64 & $>\mathcal{O}({\rm RG})$ \\
  \hline
  Determination of the semion code threshold using neural decoders \cite{PhysRevA.102.032411} & 09/2020 &SM& DP & $\sim 0.105$ & 13 & n.a. \\
  \hline
  Reinforcement learning decoders for fault-tolerant quantum computation \cite{Sweke2020Reinforcement} & 12/2020 & SC & i.i.d. bit flip \& DP & n.a. & 5 & $\gg \mathcal{O}(n)$ $\quad\quad\quad$\,\, (estimate)\\
  \hline
	Scalable Neural Decoder for Topological Surface Codes {\bf (this work)} &  01/2021 & TC & DP & 0.167 & 255 & $\mathcal{O}({\rm UF})$ \\ 
  \hline 
  \end{tabular}
  \vskip 5mm
  \caption{{\bf Literature overview} summarizing key characteristics of a selection of `conventional' decoders and ML-assisted decoders
  		for different quantum error correcting codes (QECC) including the toric code (TC), the surface code (SC), the semion code (SM), and the color code (CC).
		For each decoder, we indicate the noise model 
		(DP = depolarizing noise, i.i.d. = independent and identically distributed, CLN = circuit level noise), 
		provide the error threshold $p_{\rm th}$, the largest code distance tested $d_{\rm max}$, 
		and the algorithmic scaling in order of the total number of qubits $n$.
  		The asterisk for the threshold values for the MWPM and UF decoders indicates that these numbers have been determined in this work.
		\label{Tab:Comparisons}
  		}
\end{table*}

For the UF decoder used in our hierarchical algorithm, we have implemented a custom, open-source C++ code~\cite{UF_CYP} of the union-find algorithm and the decoder. A thorough analysis of the decoding times of the bare UF decoder and its scaling with system size for different error rates are provided in Fig.~\ref{Fig:ScalingUF}. Similar to the results in Ref.~\cite{UnionFind2017}, we find an almost perfect linear scaling of the UF decoder. Note that while the original implementation of the UF decoder has discussed only a specific error model (phase-flip with eraser errors)~\cite{UnionFind2017}, we here present results for the depolarizing noise model used throughout this manuscript. 
Data for the error threshold for this noise model for $5 \times 5$ subsystem  are shown in Fig. 2 of the main manuscript.
We additionally present finite-size scaling results with $7 \times 7$ subsystem  in Fig.~\ref{Fig:Thresholds_large} where the threshold $0.167(0)$ is obtained.

\begin{figure}[b]
	\includegraphics[width=0.99\columnwidth]{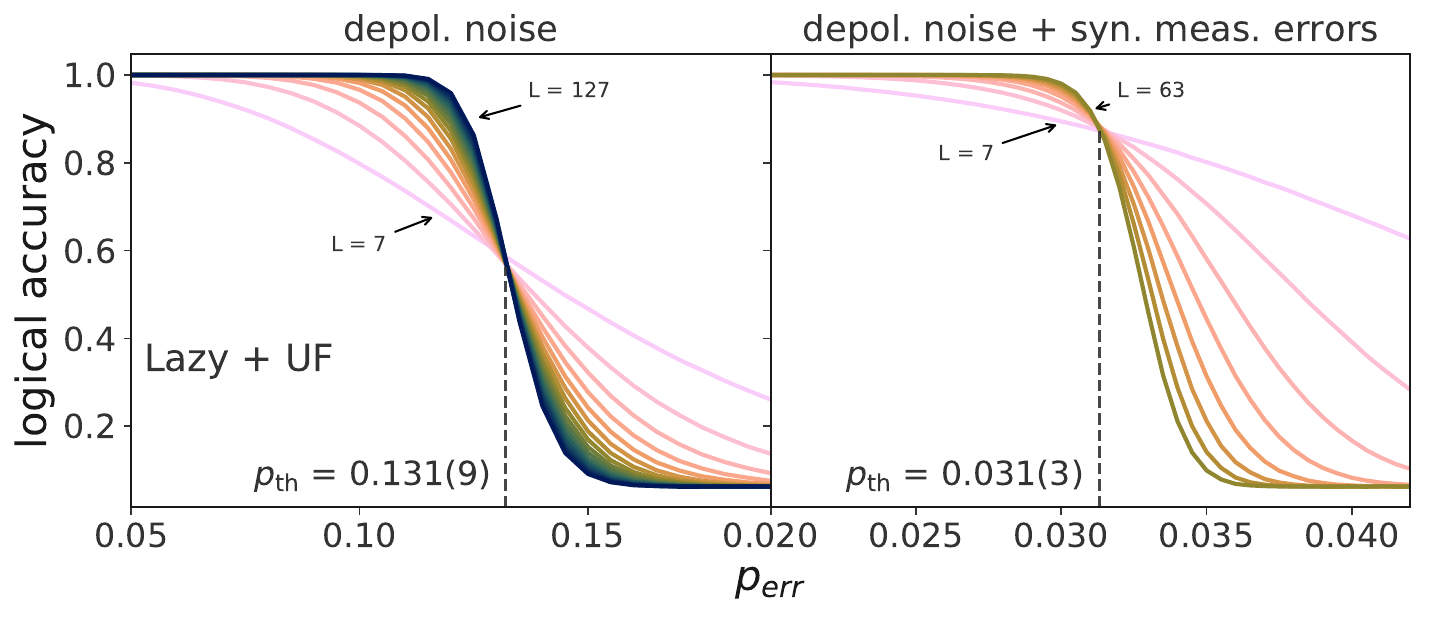}
	\caption{ {Error threshold and scaling behavior} for the {\bf lazy UF decoder} \cite{delfosse2020hardware}
			for both the depolarizing noise model (left) and the augmented noise model with additional syndrome
			measurement errors (right).
	}
	\label{Fig:LazyUF}
\end{figure}

Let us also note that when the UF decoder is used in conjunction with the ML preprocessing step,
we do {\em not} expect that the decoding time of the UF part will follow that of the bare UF decoder with a corresponding effective error rate 
(see, e.g., Fig. 3 of the main text).
This is due to the fact that the errors, which remain after the ML decoding, exhibit a significant amount of {\em long-range} corrections, and thus need more {\sl Grow} and {\sl Merge} operations than independent i.i.d.\ noise cases.
In fact, we have observed that, regardless of error rates, the UF decoder part dominates the overall decoding time in our ML+UF scheme for larger system sizes.

Finally, we present results for the error threshold measurements for the lazy UF decoder \cite{delfosse2020hardware} in Fig.~\ref{Fig:LazyUF}.


\section{Literature overview} 
\label{App:Comparision}

To allow for a detailed comparison of our ML-assisted hierarchical decoder presented in this manuscript with other decoding algorithms in the literature, in particular with regard to achievable error thresholds and algorithmic scaling, we provide a comprehensive overview of the existing literature in Table \ref{Tab:Comparisons} above.

\end{document}